 \documentclass[final,5p,times,twocolumn,compress]{elsarticle}

\usepackage{amssymb}
\usepackage{lipsum}
\usepackage[english]{babel}
\usepackage{amsmath}
\usepackage{slashed}
\usepackage{bbold}
\usepackage{cuted}
\usepackage{dirtytalk}
\usepackage{bm}
\usepackage{graphicx}
\usepackage{caption}
\usepackage{subcaption}
\usepackage{makecell}
\usepackage{gensymb}
\usepackage{float}

\journal{Physics Letters B}

\begin{document}

\begin{frontmatter}



\title{The potential of directional neutrino detection to observe neutrino spin oscillations}


\author[a]{Konstantin A. Kouzakov}
\ead{kouzakov@srd.sinp.msu.ru}
\affiliation[a]{organization={Department of Nuclear Physics and Quantum Theory of Collisions, Faculty of Physics, Lomonosov Moscow State University},
            addressline={Leninskie gory 1(2)}, 
            city={Moscow},
            postcode={119991},
            country={Russia}}

\author[b]{Fedor M. Lazarev\corref{cor1}}
\ead{lazarevfm@my.msu.ru}

\author[b]{Alexander I. Studenikin}
\ead{studenik@srd.sinp.msu.ru}
\affiliation[b]{organization={Department of Theoretical Physics, Faculty of Physics, Lomonosov Moscow State University},
            addressline={Leninskie gory 1(2)}, 
            city={Moscow},
            postcode={119991},
            country={Russia}}

\cortext[cor1]{Corresponding author}

\begin{abstract}
A nonzero neutrino magnetic moment arises already in the minimally extended Standard Model with right-handed massive Dirac neutrinos. The well-known consequence of the neutrino magnetic moment is the phenomenon of neutrino spin oscillations in a magnetic field. It can manifest itself not only as a lack in the flux of active cosmic neutrinos arriving on Earth but also as characteristic features in low-energy neutrino elastic scattering processes. Following our approach developed earlier, in this work we study the influence of arbitrary spin-flavor state of incoming neutrino on low-energy neutrino scattering off different particles in a detector. We demonstrate that superposition of left- and right-handed helicity neutrino states gives rise to an azimuthal asymmetry in the angular distribution of the recoil momenta. We present numerical calculations for elastic neutrino scattering on electrons, protons and $^{40}\!$Ar and $^{132\,}\!$Xe nuclei, demonstrating the azimuthal-asymmetry effect. Our results indicate the unique potential of directional neutrino detection to observe the neutrino spin oscillations. 
\end{abstract}




\end{frontmatter}




\section{Introduction}
\label{introduction}
Observed neutrino oscillations \cite{NobLecKajita2016,NobLecMcDonald2016} prove that neutrinos have nonzero masses. This, in turn, implies the existence of right-handed neutrinos \cite{RHN} and a nonzero neutrino magnetic moment, which in the minimal extension of the Standard Model is given by $\mu_\nu\approx3.2\times10^{-19} \mu_B\,(m_\nu/{\rm eV})$ \cite{Fujikawa1980}, where $m_\nu$ is the neutrino mass. Thus, in the general case, the neutrino spin-flavor oscillations can take place in different environments and conditions, for example, in supernovae and magnetars and in interstellar space. This phenomenon attracts a lot of attention from both theory and experiment. Since the right-handed Dirac neutrinos do not participate in weak interactions, one usually calculates and measures the fluxes of left-handed neutrinos coming to the detector. So, the lack of total active neutrino flux would indicate the spin-flavor oscillations. In the case of Majorana neutrinos, spin-flavor oscillations affect the neutrino flavor and neutrino-antineutrino flux composition.

Such spin-flavor neutrino oscillations can arise from the interaction of neutrino magnetic moments with a magnetic field. It was first proposed in Ref.~\cite{Cisneros71}. The works~\cite{Voloshin88,Akhmedov88,Lim88} predicted the effect of resonant enhancement of spin and spin-flavor oscillations during the interaction of neutrinos with a magnetic field and matter. The potential significance of this effect for oscillations of solar neutrinos and neutrinos in supernovae is discussed. A series of papers \cite{Pulido00,Akhmedov00,Akhmedov02,Akhmedov03} discusses a solution to the solar neutrino problem taking into account the effects of spin and spin-flavor oscillations as subdominant.

Thus, searches for neutrino magnetic moments and, in general, for neutrino electromagnetic (EM) properties are of particular interest~\cite{Window, AnnPhys2016,VMU2020,EMneutrinoReview2025}. Moreover, it is known that the neutrino charge radii and anapole moments are predicted to be nonzero even within the SM~\cite{Bernabeu00,Bernabeu02,Bernabeu02Erratum,Bernabeu04}.

In addition, neutrino physics is now entering the era of high-precision measurements. On the one hand, there are several next generation oscillation neutrino experiments already running or under preparation, for example, such as JUNO, Hyper-Kamiokande and DUNE. On the other hand, in the near future there will be a series of low-energy neutrino detectors, exploiting the coherent elastic neutrino-nucleus scattering process (CE$\nu$NS). Among them are COHERENT, Dresden-II, CONUS, CONNIE, $\nu$GEN, RED-100, etc. Moreover, data from COHERENT, CONUS, and Dresden-II experiments have already been used to set limits on neutrino millicharge, charge radius, and magnetic moment~\cite{NCRRadCorCorona24,Cadeddu2018,Cadeddu2018Erratum,Miranda,Cadeddu2020,CONUS2022,Dresden-II}. In this regard, an upcoming SATURNE experiment~\cite{SATURNE2024} should also be mentioned that aims to observe coherent elastic neutrino-atom scattering and to search for neutrino EM properties. 

Also dark matter detectors have been shown to be sensitive to large fluxes of solar neutrinos through elastic neutrino-electron scattering (E$\nu$ES) or CE$\nu$NS \cite{Monroe2007,Vergados2008,Strigari2009,Billard2014}. Recently, XENON and PandaX collaborations have reported the first identification of the $^8$B neutrino CE$\nu$NS event. The neutrino-induced type of background is commonly termed neutrino fog \cite{OHare2021} and challenges the experimentalists. Thus, the community has started to discuss the possibility of obtaining directional information on scattering in direct dark matter experiments \cite{Grothaus2014,OHare2015,Mayet2016,OHare2017,OHare2020}. 

There are several attempts to conduct directional measurements for neutrinos. For example, the work~\cite{Theia2020} proposed extracting directional information from Cherenkov to scintillation light ratio measurements of recoil particles. The authors of Ref.~\cite{ZekunYang2024} have presented a novel method based on waveform analysis and machine learning, which can be used for JUNO. Gas time projection chambers are also going to be a tool for providing directional information such as in the proposed \say{recoil observatory} CYGNUS project \cite{CYGNUS2024}. There are also proposals to use technologies based on solid state targets \cite{Agafonova2018,Marshall2021,Bernabei2003,Belli2020} and more exotic ones \cite{Drukier2012,Capparelli2015,Capparelli2015Erratum,OHare2022}.

In our previous work~\cite{LazarevPRD2025} we investigated the effects of all neutrino EM properties on elastic neutrino-proton scattering, where we also showed that right-handed neutrinos can scatter electromagnetically on protons on a level with the left-handed ones. In this work, we present an extended analysis of the differential cross sections of elastic neutrino scattering on electrons, protons, and nuclei in order to reveal whether the effect of superposition of incoming neutrino helicities induced by neutrino spin-flavor oscillations is strong enough to be observable in experiments. 

Thus, the structure of our work is as follows. In section \ref{RelSpin} we briefly outline the properties of the Pauli--Lubanski pseudovector, which we use as a relativistic spin and then generalize the ultrarelativistic limit of the density matrix of a Dirac particle to the case of different neutrino mass/flavor states. By such a spin-flavor density matrix we describe the incoming neutrino state. In section \ref{CS Structure} the cross sections structure for elastic neutrino scattering on an electron, a nucleon and a spinless nucleus are presented. The results of numerical calculations are presented in section \ref{NumRes} for such targets as an electron, a proton, and the $^{40}\!$Ar and $^{132\,}\!$Xe nuclei. Finally, section \ref{Conc} is devoted to our conclusions.

\section{Polarization density matrix for a Dirac particle}
\label{RelSpin}
According to Wigner, a relativistic particle is described by an irreducible representation of the Poincaré group, which contains two Casimir operators $P_\mu P^\mu$ and $W_\mu W^\mu$, where $P^\mu$ is the momentum operator (translation generator), while
\begin{equation}
    W^\mu=-\frac{1}{2}\varepsilon^{\mu\alpha\beta\gamma}M_{\alpha\beta}P_\gamma
\end{equation}
is the Pauli--Lubanski pseudovector, with $M_{\alpha\beta}$ being the Lorentz group generators. However, the Pauli--Lubanski vector has only three independent components, since it has the property $W_\mu P^\mu=0$. The eigenvalues of the Casimir operators determine the mass $m$ and the helicity modulus $j$ of the particle in accordance with
\begin{equation}
    P_\mu P^\mu=m^2,\qquad W_\mu W^\mu=-m^2j(j+1).
\end{equation}
For a Dirac particle, we have
\begin{equation}
    \begin{aligned}
        \bm{P}&=-i\bm{\nabla},\qquad P^0=(\bm{\alpha}\cdot\bm{P})+\gamma^0m,\\
        M_{\alpha\beta}&=x_\alpha P_\beta-x_\beta P_\alpha+\frac{1}{2}\sigma_{\alpha\beta},\\
        W^\mu&=\left\{\frac{1}{2}(\bm{\Sigma}\cdot\bm{P})\,,\,\frac{1}{4}\{\bm{\Sigma},P^0\}\right\}.
    \end{aligned}
\end{equation}
When acting on the plane-wave states with momentum $\bm{p}$, the Pauli--Lubanski vector is even more simple, namely $W^\mu=\frac{1}{2}\left\{(\bm{\Sigma}\cdot\bm{p})\,,\,\omega_{\bm{p}}\bm{\Sigma}\right\}$. 

Using the Pauli--Lubanski vector, one can present the density matrix of a massive Dirac particle in the form 
\begin{equation}
    \rho=\frac{1}{2}\left(\slashed{p}+m\right)\left(1-\frac{2}{m}\slashed{w}\gamma_5\right),
\end{equation}
where $w^\mu=\frac{1}{2\omega_{\bm{p}}}tr(\rho\gamma^0W^\mu)$ is the average value of the Pauli--Lubanski operator. Thus, in the rest frame, the average $4$-polarization is $2w^{\prime\mu}=\left\{0\,,\,m\bm{\zeta}\right\}$, where $\bm{\zeta}$ is the average value of the $\bm{\Sigma}$ operator in the rest frame. Passing to the laboratory system using the Lorentz transformations and introducing the longitudinal ($\parallel$) and transverse ($\perp$) polarization components with respect to the direction of particle motion, $\zeta^\parallel=(\bm{\zeta}\cdot\frac{\bm{p}}{|\bm{p}|})$ and $\bm{\zeta}^\perp=\bm{\zeta}-\zeta^\parallel\frac{\bm{p}}{|\bm{p}|}$, we obtain
\begin{equation}
    2w^\mu=\left\{(\bm{\zeta}\cdot\bm{p})\,,\,\omega_{\bm{p}}\zeta^\parallel\frac{\bm{p}}{|\bm{p}|}+m\bm{\zeta}^\perp\right\}.
\end{equation}
Then, in the ultrarelativistic limit, the density matrix can be written as
\begin{equation}
\label{Landau density matrix}
    \rho=\frac{1}{2}\slashed{p}\left(1-\zeta^\parallel\gamma_5+(\bm{\zeta}^\perp\cdot\bm{\gamma}^\perp)\gamma_5\right).
\end{equation}
Let us emphasize that the averaging of the spin operator $\bm{\Sigma}$ in the laboratory frame is given by the expression $\frac{1}{\omega_{\bm{p}}}\bm{w}=\frac{1}{2}\zeta^\parallel\frac{\bm{p}}{|\bm{p}|}+\frac{1}{2\gamma}\bm{\zeta}^\perp$, and in the ultrarelativistic limit the inverse Lorentz factor $1/\gamma$ suppresses the transverse component of the spin. Thus, the presence of the transverse component of the polarization in the rest frame in Eq.~(\ref{Landau density matrix}) is explained by the transverse component of the Pauli--Lubanski vector and is interpreted as a quantity characterizing the superposition of left- and right-handed helicity states of the particle in the ultrarelativistic limit.

In our previous work \cite{LazarevPRD2025} we obtained the generalization to different mass/flavor ultrarelativistic neutrino states
\begin{equation}
\label{Landau neutrino density matrix}
    \rho_{ij}=\frac{1}{2}\slashed{k}\left(\tilde\rho_{ij}-\zeta^\parallel_{ij}\gamma_5+(\bm{\zeta}^\perp_{ij}\cdot\bm{\gamma}_\perp)\gamma_5\right),
\end{equation}
where $\tilde\rho_{ij}=\frac{1}{2E_\nu}tr(\rho_{ij}\gamma^0)$ is a reduced density matrix in the neutrino mass space. $\frac{1}{2}\zeta^\parallel_{ij}$ and ${\bm\zeta}^\perp_{ij}=\zeta^x_{ij}\bm{e}_x+\zeta^y_{ij}\bm{e}_y$, with $\{{\bm e}_x,{\bm e}_y,{\bm k}/E_\nu\}$ forming a 3-vector basis, are the matrices of the longitudinal and transverse components, with respect to the neutrino momentum $\bm{k}$, corresponding to the diagonal ($i=j$) and transition ($i\neq j$) values of the spin operator $\bm{\Sigma}$ averaged over spin states of the neutrino in its rest frame. If there is a state with zero neutrino mass, the diagonal and transition ${\zeta}^\perp$ components involving this state vanish. Also we note that the density matrix~(\ref{Landau neutrino density matrix}) can be determined taking into account that in the ultrarelativistic limit the components of the spin-flavor density matrix in the mass basis typically employed in neutrino oscillation calculations are related to the components $\frac{1}{4E^2_\nu}(u^{(\nu_i)}_{{\bm k},h'})^\dagger\rho_{ij}\gamma^0u^{(\nu_j)}_{{\bm k},h}$, where $u^{(\nu_i)}_{k,r}$ stands for the bispinor amplitude of the $\nu_i$ state with momentum ${\bm k}$ and helicity $h$. Such a density matrix, acting on the column vectors in the basis of neutrino states $\{(\nu^R_1,\nu^R_2,\nu^R_3),(\nu^L_1,\nu^L_2,\nu^L_3)\}$, can be presented as
\begin{equation}
\label{Oscill density matrix}
    \rho=\frac{1}{2}\begin{pmatrix}
    \tilde\rho+\zeta^\parallel & \zeta^x-i\zeta^y\\
    \zeta^x+i\zeta^y& \tilde\rho+\zeta^\parallel
    \end{pmatrix}=\frac{1}{2}\bigg(\tilde\rho\mathbb{1}+(\bm{\zeta}\cdot\hat{\bm{\sigma}})\bigg),
\end{equation}
where $\mathbb{1}$ and the Pauli matrices $\hat{\bm{\sigma}}$ are assumed to be the block matrices with their components being matrices acting on the mass states.

%
%
\section{Differential cross sections}
\label{CS Structure}
In the case of directional neutrino measurements one studies the angular distribution of recoil particles. Assuming the target particle is initially at rest, the angular differential cross section for elastic neutrino scattering has the general structure
\begin{equation}
    \frac{d\sigma}{d\Omega}=A(\theta)+B_s(\theta)\sin{\varphi}+B_c(\theta)\cos{\varphi},
\end{equation}
where $\theta$ and $\varphi$ are the polar and azimuthal angles of the recoil particle momentum. 

The analysis of specific contributions is more transparent in terms of the recoil energy $T$ (energy transfer), rather than the polar angle $\theta$. They are related as follows
\begin{equation}
\label{ThetaTRelatio}
    \begin{aligned}
        \cos\theta&=\sqrt{\frac{T}{T+2m_t}}\frac{E_\nu+m_t}{E_\nu},
    \end{aligned}
\end{equation}
where $m_t$ is the recoil particle mass ($m_t=m_e$ or $m_N$ for an electron or a nucleon, respectively, and $m_t=M$ for a nucleus). Introducing the double differential cross section
\begin{equation}
\label{DDCS}
\frac{d^2\sigma}{dTd\varphi}=\tilde{A}(T)+\tilde{B}_s(T)\sin{\varphi}+\tilde{B}_c(T)\cos{\varphi}
\end{equation}
and using the relation (\ref{ThetaTRelatio}), we find that
\begin{equation}
\label{ThetaTRelatio1}
    \begin{aligned}
      X(\theta)=\frac{E_\nu\sqrt{T(T+2m_t)^{3}}}{m_t(E_\nu+m_t)}\,\tilde{X}(T), \qquad (X=A,B_s,B_c).  
    \end{aligned}
\end{equation}

There are four distinct terms in the cross section formulas. The first pair does not depend on the azimuthal angle $\varphi$. One of these terms is determined by weak interaction, and the other one by the square of neutrino magnetic moment. The second pair is proportional to the neutrino magnetic moment and transverse spin polarization. Since the full expressions for the neutrino-nucleon scattering are cumbersome (see \ref{FullFormulas}) and cross sections depend on combinations such as $\frac{T}{E_\nu}$ and $\frac{T}{m_N}$, which are less than unity in the kinematic regime considered, below we write out only the leading terms with respect to the energy transfer $T$ for each contribution we are interested in: 
%
\begin{equation}
\label{AzAsCSp}
    \begin{aligned}
        \tilde{A}(T)&=\frac{G_F^2m_N}{2\pi^2}\left[C_V^N+C_A^N+(C_A^N-C_V^N)\frac{m_NT}{2E_\nu^2}\right]+\\
        &+\frac{\alpha^2|\mu_\nu|^2}{2m_e^2E_\nu}\left[\frac{E_\nu}{T}(F^N_Q)^2\right],\\
        \tilde{B}_s(T)&=\frac{G_F\alpha}{m_e\pi}\sqrt{\frac{m_N}{T}}\textrm{Im}K_{\mu_\nu,Z^0}F^N_1F^N_Q+\\
        &+\frac{4\sqrt{2}\alpha^2}{m_e}\sqrt{\frac{m_N}{T}}\textrm{Im}K_{\mu_\nu,a_\nu}(F^N_Q)^2,\\
        \tilde{B}_c(T)&=-\frac{G_F\alpha}{m_e\pi}\sqrt{\frac{m_N}{T}}\textrm{Re}K_{\mu_\nu,Z^0}F^N_1F^N_Q-\\
        &-\frac{4\sqrt{2}\alpha^2}{m_e}\sqrt{\frac{m_N}{T}}\textrm{Re}K_{\mu_\nu,a_\nu}(F^N_Q)^2,\\
    \end{aligned}
\end{equation}
%
with 
\begin{equation}
\label{AzAsCSp_Coeff}
\begin{aligned}
     &C^N_V=\frac12Tr\left[\left(-F_1^N\mathbb{1}+F_Q^NQ^L\right)^2\tilde\rho\right],\\
     &C^N_A=\frac12Tr\left[\left(G^N_A\mathbb{1}\right)^2\tilde\rho\right],\\   
     &Q^L=\frac{2\sqrt{2}\pi\alpha}{G_F}\left(\frac16\langle r_\nu^2\rangle-a_\nu\right), \quad  |\mu_\nu|^2=Tr\left[\mu_\nu^2\tilde\rho\right],\\
     &K_{\mu_\nu,Z^0}=Tr\left[\mu_\nu\kappa^\dagger\right], \quad K_{\mu_\nu,a_\nu}=Tr\left[a_\nu \mu_\nu\kappa^\dagger\right],\\
     &\kappa_{ij}=\frac12(\zeta^x_{ij}+i\zeta^y_{ij}),
\end{aligned}
\end{equation}
where trace is taken over the matrices acting on the mass/flavor neutrino states. Definitions of the nucleon form factors $F^N_1$, $F_Q^N$, and $G_A^N$ ($N=p,n$) see in \ref{FullFormulas}. 

In the case of neutrino scattering on a spin-zero nucleus, the full expressions are
\begin{equation}
\label{AzAsCSNucl}
    \begin{aligned}
        \tilde{A}(T)&=\frac{G_F^2M}{2\pi^2}\left(1-\frac{T}{E_\nu}-\frac{MT}{2E_\nu^2}\right)C^{Nucl}_V+\\
        &+\frac{\alpha^2|\mu_\nu|^2}{2m_e^2}\left(\frac{1}{T}-\frac{1}{E_\nu}+\frac{T^2}{4E_\nu^2}\right)\mathcal{F}_Q^2,\\
        \tilde{B}_s(T)&=\frac{\alpha G_F}{\pi m_e}\sqrt{MT\left(1-\frac{T}{E_\nu}-\frac{TM}{2E_\nu^2}\right)}\left(\frac{1}{T}-\frac{1}{2E_\nu}\right)\cdot\\
        &\cdot\Big[\textrm{Im}K_{\mu_\nu,Z^0}\mathcal{F}_Q\mathcal{F}_1+\frac{4\sqrt{2}\pi\alpha}{G_F}\textrm{Im}K_{\mu_\nu,a_\nu}\mathcal{F}_Q^2\Big]\\
        \tilde{B}_c(T)&=-\frac{\alpha G_F}{\pi m_e}\sqrt{MT\left(1-\frac{T}{E_\nu}-\frac{TM}{2E_\nu^2}\right)}\left(\frac{1}{T}-\frac{1}{2E_\nu}\right)\cdot\\
        &\cdot\Big[\textrm{Re}K_{\mu_\nu,Z^0}\mathcal{F}_Q\mathcal{F}_1+\frac{4\sqrt{2}\pi\alpha}{G_F}\textrm{Re}K_{\mu_\nu,a_\nu}\mathcal{F}_Q^2\Big],
    \end{aligned}
\end{equation}
with
\begin{equation}
\label{AzAsCSNucl_Coeff}
\begin{aligned}
     &C^{Nucl}_V=\frac{1}{2}Tr\left[\left(-\mathcal{F}_1\mathbb{1}+\mathcal{F}_QQ^L\right)^2\tilde\rho\right],\\
     &\mathcal{F}_Q(\bm{q})=\int d^3re^{i(\bm{q}\cdot\bm{r})}\langle 0|\sum_{a=1}^{A}Q_a\delta(\bm{r}-\bm{r}_a)|0\rangle,\\
     &\mathcal{F}_1(\bm{q})=\int d^3re^{i(\bm{q}\cdot\bm{r})}\langle 0|\sum_{a=1}^{A}g_V^a\delta(\bm{r}-\bm{r}_a)|0\rangle,
\end{aligned}
\end{equation}
where $|0\rangle$ is the nuclear ground state, $A$ is a mass number, $Q_a$ is an electric charge of the nucleon ($1$ for protons and $0$ for neutrons), and $g_V^a$ is a weak vector-coupling constant, which is $g_V^p=F_1^p(q=0)$ for protons and $g_V^n=F^n_1(q=0)$ for neutrons.

It can be seen that the leading terms for each contribution have a similar structure. Specifically, there is a linear term with respect to energy transfer $T$ on the scale $\frac{G_F^2m_t}{2\pi^2}C_w$, where the factor $C_w$ is determined by the weak-interaction properties of the target particle ($C_V^N$ and $C_A^N$ in the nucleon case and $C_V^{Nucl}$ in the nucleus case). In addition, there is the behavior $\frac{E_\nu}{T}$ on the scale $\frac{\alpha^2|\mu_\nu|^2}{2m_e^2E_\nu}Q^2$, where $Q$ is the electric charge of the target particle. Finally, the azimuthal asymmetry behaves as $\sqrt{\frac{m_t}{T}}$ and is on the scales $\frac{G_F\alpha}{m_e\pi}\mu_\nu Q_wQ$ and $\frac{4\sqrt{2}\alpha^2}{m_e}a_\nu\mu_\nu Q^2$, with $Q_w$ being the weak charge of the target particle.

\begin{table*}[h]
    \centering
    \begin{tabular}{c||c|c|c|c}
        Particle & $m$ & $Q$ & $Q_w$ & $C_w$ \\ \hline
        $e^-$ & 511 keV & -1 & \makecell{$g_V^e=\mp\frac{1}{2}+2s^2_W=-0.038(0.962)$} & $0.13(0.57)$ \\ \hline
        p & 938 MeV & 1 & \makecell{$g_V^p=\frac{1}{2}-2s^2_W=0.038$} & $0.13$ \\ \hline
        n & 940 Mev & 0 & \makecell{$g_V^n=-0.5$} & $0.25$ \\ \hline
        $^{40}\!$Ar & 37.2 GeV & 18 & \makecell{$\mathcal{F}_1=18g_V^p+22g_V^n=$\\$=-10.324$} & $53.29$ \\ \hline
        $^{132\,}\!$Xe & 122.9 GeV & 54 & \makecell{$\mathcal{F}_1=54g_V^p+78g_V^n=$\\$=-36.971$} & $683.43$ \\
    \end{tabular}
    \caption{Mass and weak and electric charges of target particles}
    \label{EWProp}
\end{table*}
\begin{table*}[h]
    \centering
    \begin{tabular}{c||c|c|c|c}
        Particle & $\frac{G_F^2m}{2\pi^2}C_w$ & $\frac{\alpha^2|\mu_\nu|^2}{2m_e^2E_\nu}Q^2$ & $\frac{G_F\alpha}{m_e\pi}\mu_\nu Q_wQ$ & $\frac{4\sqrt{2}\alpha^2}{m_e}a_\nu\mu_\nu Q^2$ \\ \hline
        $e^-$ & \makecell{$4.5\times10^{-25}$\\$(2.0\times10^{-24})$} & $2.3\times10^{-27}$ & \makecell{$3.0\times10^{-26}$\\$(7.7\times10^{-25})$} & $4.7\times10^{-26}$ \\ \hline
        p & $8.2\times10^{-22}$ & $2.3\times10^{-27}$ & $3.0\times10^{-26}$ & $4.7\times10^{-26}$ \\ \hline
        n & $1.6\times10^{-21}$ & $0$ & $0$ & $0$ \\ \hline
        $^{40}\!$Ar & $1.4\times10^{-17}$ & $7.4\times10^{-25}$ & $1.5\times10^{-22}$ & $1.5\times10^{-23}$ \\ \hline
        $^{132\,}\!$Xe & $5.8\times10^{-16}$ & $6.7\times10^{-24}$ & $1.6\times10^{-21}$ & $1.4\times10^{-22}$ \\
    \end{tabular}
    \caption{Scales of different contributions}
    \label{ContibScale}
\end{table*}
In table \ref{EWProp} the necessary properties of target particles are presented, while table \ref{ContibScale} presents numerical estimates for the aforementioned scales of specific contributions. In these estimates we take neutrino diagonal magnetic moments at the level of $10^{-11}\mu_B$ according to the limits~\cite{LUX23}
\begin{equation}
\label{mu_nu_diag}
\begin{aligned}   
|\mu_{ee}|&<1.5\times10^{-11}\mu_B,\\
|\mu_{\mu\mu}|&<2.3\times10^{-11}\mu_B,\\
|\mu_{\tau\tau}|&<2.1\times10^{-11}\mu_B,
\end{aligned}
\end{equation}
and neutrino charge radii and anapole moments to be given by their SM values\footnote{Our definition of the SM neutrino charge radii differs in sign from that in Refs.~\cite{Bernabeu00,Bernabeu02,Bernabeu04} for the reasons explained in Ref.~\cite{Cadeddu2018}.}~\cite{Bernabeu00,Bernabeu02,Bernabeu04}
\begin{equation}
\label{SM CR values}
    \begin{aligned}
    \langle r_{ee}^2\rangle_{SM}&=-4.1\times10^{-33}\,\textrm{cm}^2,\\
    \langle r_{\mu\mu}^2\rangle_{SM}&=-2.4\times10^{-33}\,\textrm{cm}^2,\\
    \langle r_{\tau\tau}^2\rangle_{SM}&=-1.5\times10^{-33}\,\textrm{cm}^2.
    \end{aligned}
\end{equation}
It is pointed out in Ref.~\cite{Bernabeu02} that the SM also predicts nonzero values for the neutrino diagonal in the flavor basis anapole moments,
\begin{equation}
\label{SM vAnapole}
    a^{SM}_{\ell\ell}=-\frac{1}{6}\langle r^2_{\ell\ell}\rangle_{SM}.
\end{equation}
The unpolarized and transversely polarized spin neutrino states are considered, so that $\zeta^\parallel_{ij}\equiv0$. Finally, the neutrino energy is $E_\nu=10$ MeV, which is characteristic of neutrinos from a core-collapse supernova explosion. 

It is already clear that the effect of azimuthal asymmetry can be appreciable in the low-energy domain, where the squared magnetic moment contribution becomes dominant instead of purely weak-interaction one. There is a growth in the contribution of weak interactions for nuclei due to the coherence scattering regime. Other contributions remain more or less at the same level, with the exception of the contribution in the third column. That contribution is of particular interest because it is proportional to the product of the weak and electric charges. In the case of a proton, we have a weak charge close to zero, and in the case of a neutron, the electric charge is zero, but for a nucleus, both weak and electric charges are nonzero, which causes the discussed contribution to grow.

By selecting targets with large weak and electric charges, we lose in target recoil energies due to the growth of its mass. Accordingly, we can take a different approach and consider the lightest target particles, i.e. electrons, minimizing the contribution of weak interactions. For them, we also obtained analytical expressions and performed numerical modeling. We point out that general formulas for scattering on electrons can be obtained by replacing nucleon form factors with electron ones. However, we must also take into account the additional contribution of charged currents in the case of scattering of an electron neutrino on electrons. Namely $\delta_{ij}F_1^N(q^2)\rightarrow(\hat{g}_V^e)_{ij}=g_V^e\delta_{ij}+U^*_{ei}U_{ej}$ and $\delta_{ij}G_A^N(q^2)\rightarrow(\hat{g}_A^e)_{ij}=g_A^e\delta_{ij}+U^*_{ei}U_{ej}$
, where $U$ is the neutrino mixing matrix, and the weak charge coupling constants are $g_V^e=-\frac12+2s^2_W$ and $g_A^e=-\frac12$. We present below for comparison contributions similar to nucleon and nuclear cases, however, noting that they are no longer necessarily leading, since the energy of the incident neutrino is greater than the mass of the electron, and accordingly the energy transfer can also be comparable with the electron mass:
\begin{equation}
\label{AzAsCSe}
    \begin{aligned}
        \tilde{A}(T)=&\frac{G_F^2m_e}{2\pi^2}\left[C_V^{e^-}+C_A^{e^-}+\left(C_A^{e^-}-C_V^{e^-}\right)\frac{m_eT}{2E_\nu^2}\right]+\\
        +&\frac{\alpha^2|\mu_\nu|^2}{2m_e^2E_\nu}\left[\frac{E_\nu}{T}\right],\\
        \tilde{B}_s(T)=&-\frac{G_F\alpha}{\pi\sqrt{m_eT}}\textrm{Im}K_{\mu_\nu,Z^0}g_V^e-\\
        -&\frac{G_F\alpha}{\pi\sqrt{m_eT}}\textrm{Im}K_{\mu_\nu,W}+\frac{4\sqrt{2}\alpha^2}{\sqrt{m_eT}}\textrm{Im}K_{\mu_\nu,a_\nu},\\
       \tilde{B}_c(T)=&\frac{G_F\alpha}{\pi\sqrt{m_eT}}\textrm{Re}K_{\mu_\nu,Z^0}g_V^e+\\
        +&\frac{G_F\alpha}{\pi\sqrt{m_eT}}\textrm{Re}K_{\mu_\nu,W}-\frac{4\sqrt{2}\alpha^2}{\sqrt{m_eT}}\textrm{Re}K_{\mu_\nu,a_\nu},\\
    \end{aligned}
\end{equation}
\begin{equation}
\label{AzAsCSe_Coeff}
\begin{aligned}
     &C_V^{e^-}=\frac12Tr\left[\left(-\hat{g}_V^e-Q^L\right)^2\tilde\rho\right],\quad (\hat{g}_V^e)_{ij}=g_V^e\delta_{ij}+U^*_{ei}U_{ej}\\
     &C_A^{e^-}=\frac12Tr\left[\left(\hat{g}_A^e\right)^2\tilde\rho\right],\quad (\hat{g}_A^e)_{ij}=g_A^e\delta_{ij}+U^*_{ei}U_{ej}\\
     &K_{\mu_\nu,W}=Tr\left[(\hat{g}_V^e-g_v^e\mathbb{1})\mu_\nu\kappa^\dagger\right].
\end{aligned}
\end{equation}
For $\nu_ee^-$-scattering, additional contributions from the charged current arise, and some of them (those proportional to $K_{\mu_\nu,W}$) enhance the azimuthal asymmetry of the cross section. In  table \ref{ContibScale}, the values in brackets correspond to the $\nu_ee^-$-scattering case. 

\section{Numerical illustration}
\label{NumRes}
We performed numerical calculations to demonstrate the effect of azimuthal asymmetry of differential cross sections for neutrino scattering on a proton, the $^{40}\!$Ar and $^{132\,}\!$Xe nuclei, and an electron. The numerical results were obtained on the basis of exact expressions for the cross sections, i.e. without involving only the leading contributions discussed above.

\begin{figure*}
    \centering
        \begin{subfigure}{0.35\textwidth} 
        \centering
        \includegraphics[width=\textwidth]{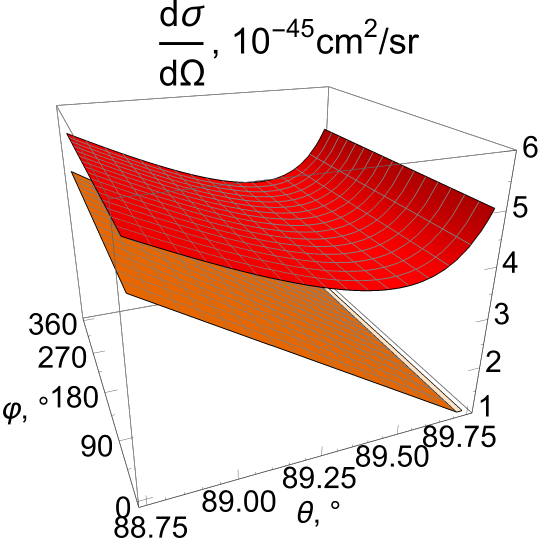} 
        \end{subfigure}
    \begin{subfigure}{0.35\textwidth}
        \centering
        \includegraphics[width=\textwidth]{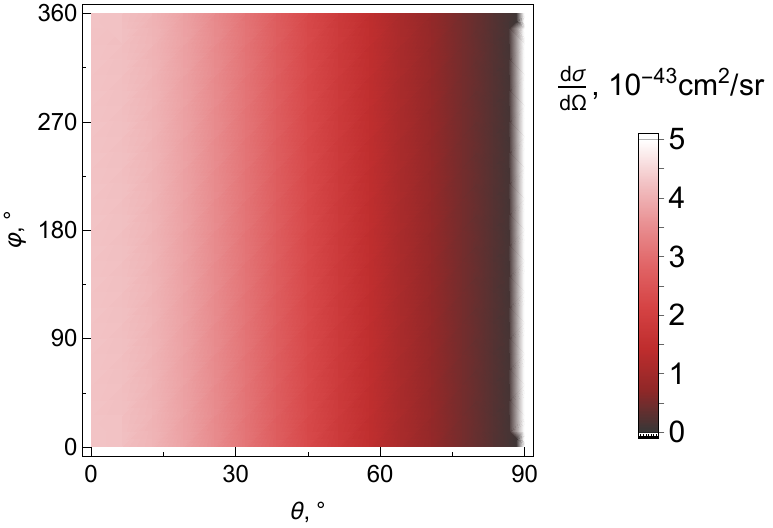}
        \end{subfigure}
    \hfill 
    \begin{subfigure}{0.35\textwidth} 
        \centering
        \includegraphics[width=\textwidth]{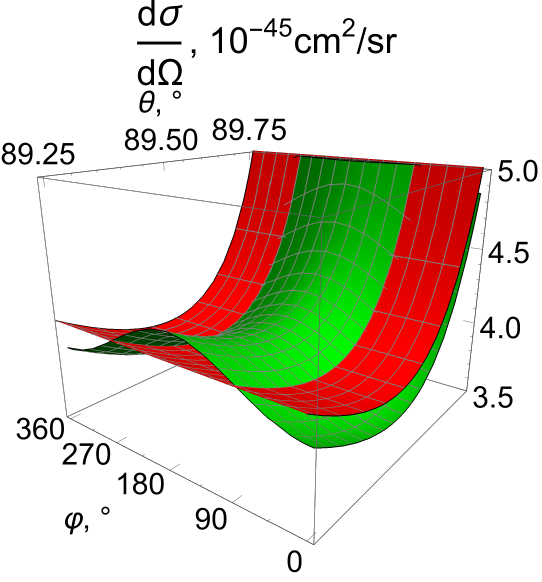} 
        \end{subfigure}
    \begin{subfigure}{0.35\textwidth}
        \centering
        \includegraphics[width=\textwidth]{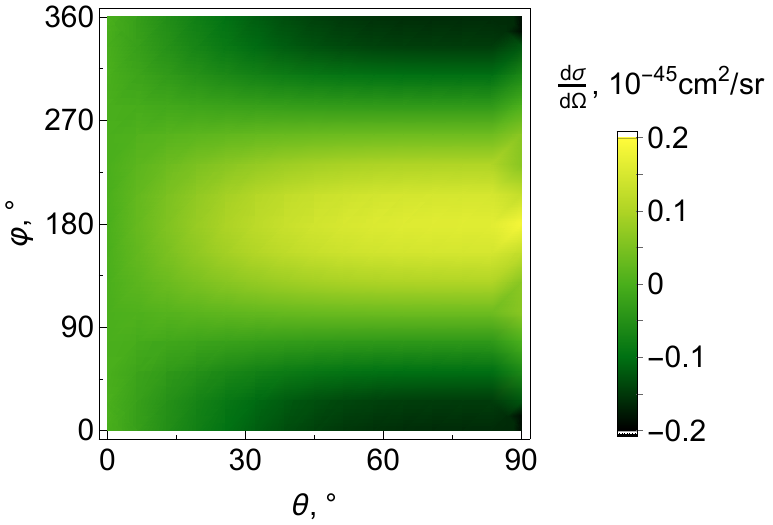}
        \end{subfigure}
    \begin{subfigure}{0.35\textwidth} 
        \centering
        \includegraphics[width=\textwidth]{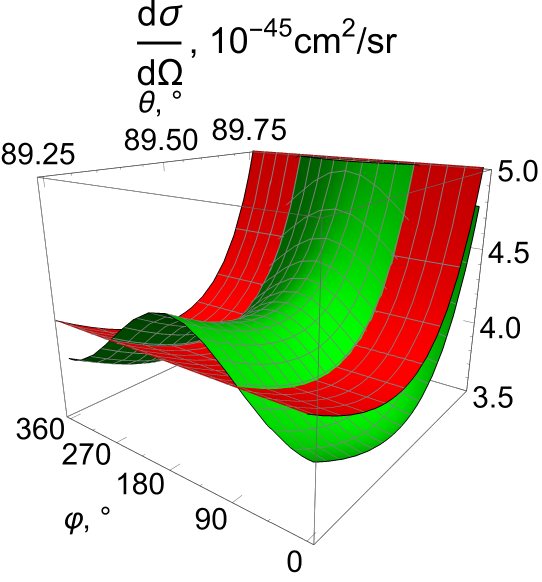} 
        \end{subfigure}
    \begin{subfigure}{0.35\textwidth}
        \centering
        \includegraphics[width=\textwidth]{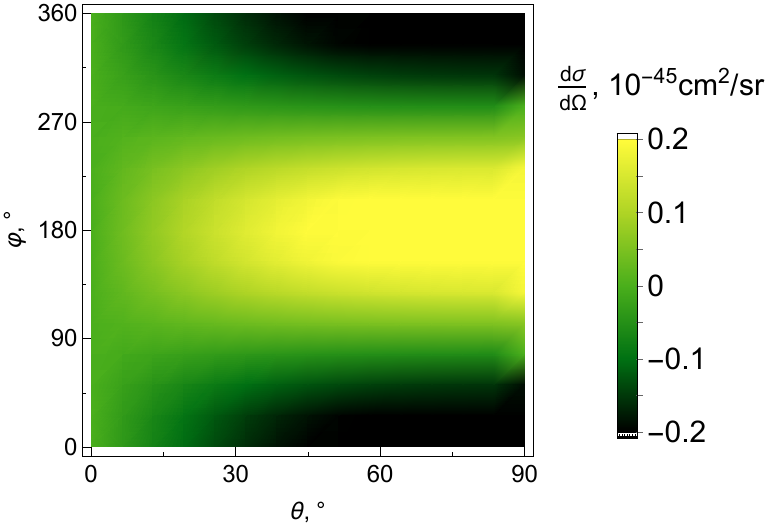}
        \end{subfigure}
    \center{\includegraphics[width=0.65\linewidth]{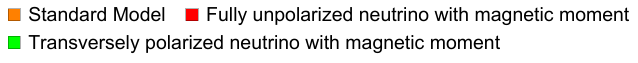}}
    \caption{The angular differential cross sections for elastic neutrino-proton scattering for different incident neutrino spin states. In the first row, the 3D plot presents the cross sections for fully unpolarized neutrinos (i) within the SM limits and (ii) taking into account the neutrino magnetic moments of $10^{-11}\mu_B$, while the 2D density plot shows only the latter cross section. In the second row, the 3D plot compares the cross sections taking into account the neutrino magnetic moments of $10^{-11}\mu_B$ in the cases of fully unpolarized and transversely polarized neutrinos, while the density plot shows the difference between these cross sections. Finally, the third row is the same as the second, but taking into account the influence of the SM values of the neutrino charge radii and anapole moments.}
\label{MMtp}
\end{figure*}

Figure~\ref{MMtp} illustrates the effect of the neutrino transverse spin polarization on the cross section differential with respect to the solid angle of the recoil proton. Note that only in the presence of a transverse component of the neutrino spin polarization the cross section does depend on the azimuthal angle of the recoil momentum. We assume that the incident neutrino is completely transversely polarized with respect to its momentum and compare this case with that when the neutrino is completely unpolarized. As follows from Eqs.~(\ref{AzAsCSp}) and (\ref{AzAsCSp_Coeff}), the contribution of the transverse spin polarization of the neutrino to the cross section is proportional to neutrino magnetic moments, so that it completely vanishes if they have zero values. The effect occurs on such small scales that the contributions from the SM values of the neutrino charge radii and anapole moments also become noticeable, and, moreover, they tend to enhance the effect under discussion.

\begin{figure*}
    \centering
        \begin{subfigure}{0.35\textwidth} 
        \centering
        \includegraphics[width=\textwidth]{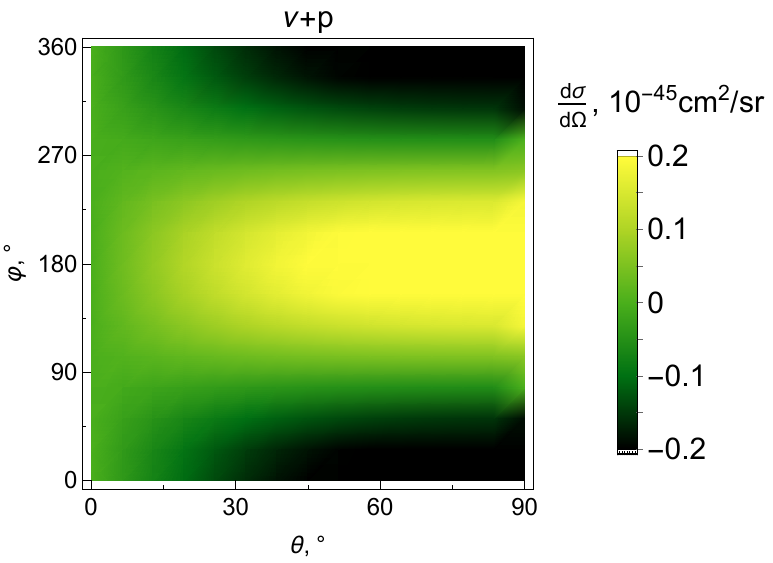} 
        \end{subfigure}
    \begin{subfigure}{0.35\textwidth}
        \centering
        \includegraphics[width=\textwidth]{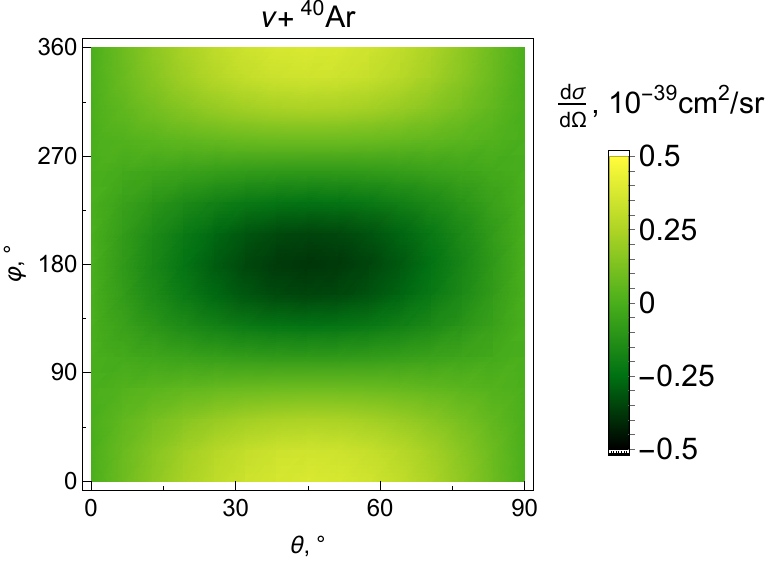}
        \end{subfigure}
    \hfill 
    \begin{subfigure}{0.35\textwidth} 
        \centering
        \includegraphics[width=\textwidth]{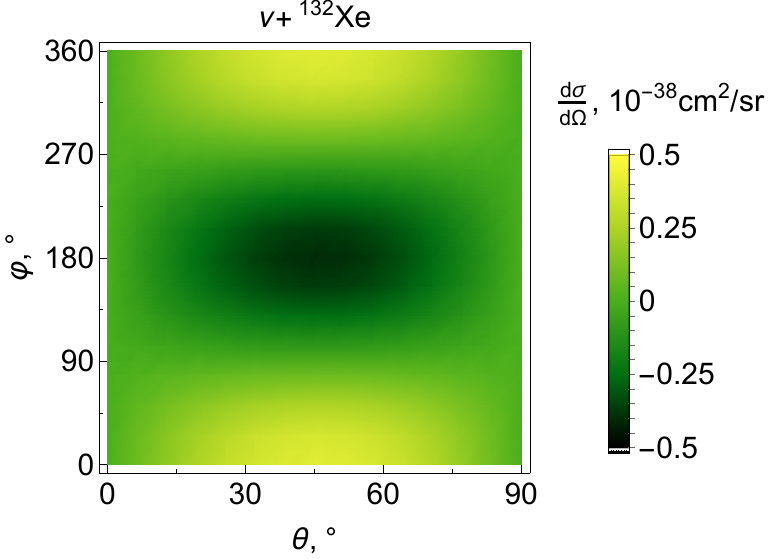} 
        \end{subfigure}
    \caption{The angular differential cross sections for elastic neutrino scattering on a proton and the $^{40}\!$Ar and $^{132\,}\!$Xe nuclei in the case of transverse spin polarization of neutrinos. A neutrino magnetic moment of $10^{-11}\mu_B$ and the SM values of neutrino charge radii and anapole moments are taken into account.}
\label{MMtNucl}
\end{figure*}

Figure \ref{MMtNucl} compares the effects of azimuthal asymmetry of the cross section for neutrino scattering on a proton and on the $^{40}\!$Ar and $^{132\,}\!$Xe nuclei. Note that in the case of a proton, the region where this effect is clearly visible is at a polar angle $\theta$ close to $90\degree$, i.e. in the region where the contribution of weak interactions decreases and the contribution of the square of the neutrino magnetic moment begins to dominate. That is, the effect is mostly visible against the background of a minimum of the cross section as a function of the polar angle. Note that the contribution of weak interactions decreases with increasing the proton polar angle in neutrino-proton scattering. In the case of scattering on nuclei, the contribution of weak interactions has a maximum, and the region close to $\theta\approx90\degree$, where the cross section begins to increase again, becomes very narrow, and therefore the effect of azimuthal asymmetry of the cross section is visible precisely against the background of the maximum.

\begin{figure*}
    \centering
        \begin{subfigure}{0.35\textwidth} 
        \centering
        \includegraphics[width=\textwidth]{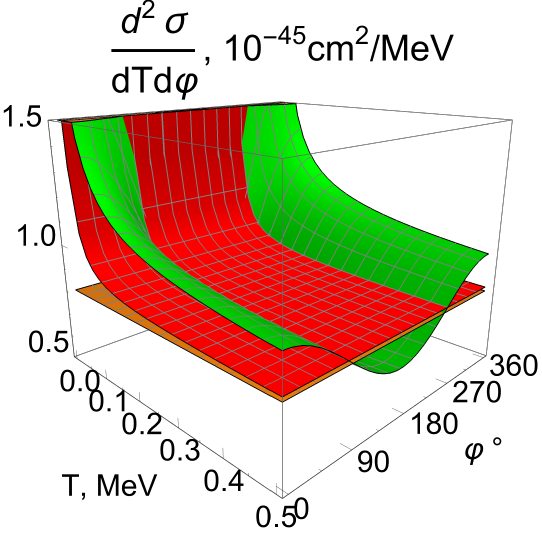} 
        \end{subfigure}
    \begin{subfigure}{0.35\textwidth}
        \centering
        \includegraphics[width=\textwidth]{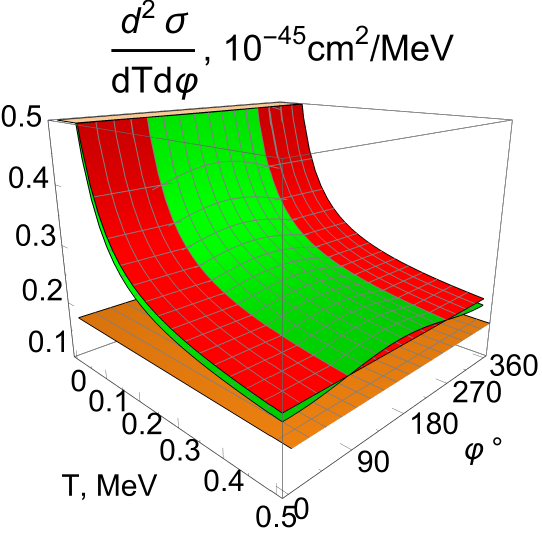}
        \end{subfigure}
    \center{\includegraphics[width=0.65\linewidth]{legendMMt.pdf}}
    \caption{The double differential cross section~(\ref{DDCS}) for elastic neutrino-electron scattering for different incident neutrino spin states. On the left is the $\nu_e$ case, and on the right is the $\nu_{\mu,\tau}$ case.}
\label{MMte}
\end{figure*}

Figure \ref{MMte} shows the double differential cross section~(\ref{DDCS}) for elastic neutrino-electron scattering. Note that in the electron case the effect of azimuthal asymmetry turns out to be maximal and can be seen even at recoil energies comparable to the electron mass.

\section{Summary and conclusions}
\label{Conc}
The effect of azimuthal asymmetry in the angular distributions of recoil particles in elastic neutrino scattering processes, which can arise for nonzero neutrino magnetic moments along with certain neutrino spin polarizations, is described in detail and proposed for experimental observation with the method of directional neutrino detection. Namely, from the point of view of the laboratory frame, in which the neutrino moves ultrarelativistically, the neutrino spin states are superpositions of left- and right-handed neutrinos. On the other hand, since the neutrino has a nonzero mass, these states can be obtained by setting the neutrino spin in the rest frame along a direction that does not coincide with the direction of the boost for the transition to the laboratory frame.

The scattering of such neutrino states on various target particles is analyzed in the present work. Despite its smallness, the effect appears to be nonzero in all the cases considered: for scattering on a proton, the $^{40}\!$Ar and $^{132\,}\!$Xe nuclei, and an electron. Note that the effect manifests itself in different ranges of target particle recoil angles, depending on the particle itself. Also, the case of $\nu_ee^-$-scattering demonstrates a fairly strong effect, which is consistent with the fact that the most stringent laboratory constraints on the neutrino magnetic moment are obtained from elastic neutrino-electron scattering experiments. Observation of the azimuthal asymmetry effect in the experiment will provide evidence of both the nonzero neutrino magnetic moment and the phenomenon of neutrino spin oscillations.

\section*{Acknowledgments}
This work is supported by the Russian Science Foundation (project No. 24-12-00084).

\appendix

\section{Full expressions for differential cross sections of elastic neutrino-nucleon and neutrino-electron scattering}
\label{FullFormulas}
The formulas for differential cross sections of elastic neutrino-nucleon scattering, which account for all neutrino electromagnetic properties, were derived in our previous work \cite{LazarevPRD2025}. Here we present the full expressions corresponding to the specific case considered in this study:
%
\begin{equation}
\label{FullCSp}
    \begin{aligned}
        \tilde{A}(T)&=\frac{G^2_Fm_N}{4\pi^2}\Bigg[\left(C^N_V-2{\rm Re}\,C^N_{V\&A}+C^N_A\right)+\\
        &+\left(C^N_V+2{\rm Re}\,C^N_{V\&A}+C^N_A\right)\left(1-\frac{T}{E_\nu}\right)^2\\
     &+\left(C^N_A-C^N_V\right)\frac{m_NT}{E^2_\nu}+C^N_M\frac{T}{2m_N}\left(2+\frac{m_NT}{E_\nu^2}-\frac{2T}{E_\nu}\right)+\\
     &+2\frac{T}{E_\nu}{\rm Re}\,C^N_{A\&M}\left(2-\frac{T}{E_\nu}\right)-2{\rm Re}\,C^N_{V\&M}\frac{T^2}{E^2_\nu}\Bigg]+\\
     &+\frac{\alpha^2}{2m^2_e}|\mu_\nu|^2\Bigg[\left(\frac{1}{T}-\frac{1}{E_\nu}\right)(F^N_Q)^2-\frac{T}{2E^2_\nu}F^N_QF^N_M+\\
     &+\frac{(F^N_M)^2}{8m_N}\left[\left(2-\frac{T}{E_\nu}\right)^2-\frac{2m_NT}{E^2_\nu}\right]\Bigg],\\
        \tilde{B}_s(T)&=\frac{G_F\alpha}{m_e\pi}\sqrt{\frac{m_N}{T}\left(1-\frac{T}{E_\nu}-\frac{m_NT}{2E_\nu^2}\right)}\Bigg\{\Bigg[\left(1-\frac{T}{2E_\nu}\right)\times\\
        &\times\left(F^N_1F^N_Q+\frac{T}{2m_N}F^N_2F^N_M\right)+\frac{T}{2E_\nu}G^N_A\left(F^N_Q+F^N_M\right)\Bigg]\textrm{Im}K_{\mu_\nu,Z^0}+\\
        &+\frac{4\sqrt2\pi\alpha}{G_F}\left(1-\frac{T}{2E_\nu}\right)\left((F^N_Q)^2+\frac{T}{2m_N}(F^N_M)^2\right)\textrm{Im}K_{\mu_\nu,a_\nu}\Bigg\},\\
        \tilde{B}_c(T)&=-\frac{G_F\alpha}{m_e\pi}\sqrt{\frac{m_N}{T}\left(1-\frac{T}{E_\nu}-\frac{m_NT}{2E_\nu^2}\right)}\Bigg\{\Bigg[\left(1-\frac{T}{2E_\nu}\right)\times\\
        &\times\left(F^N_1F^N_Q+\frac{T}{2m_N}F^N_2F^N_M\right)+\frac{T}{2E_\nu}G^N_A\left(F^N_Q+F^N_M\right)\Bigg]\textrm{Re}K_{\mu_\nu,Z^0}-\\
        &-\frac{4\sqrt2\pi\alpha}{G_F}\left(1-\frac{T}{2E_\nu}\right)\left((F^N_Q)^2+\frac{T}{2m_N}(F^N_M)^2\right)\textrm{Re}K_{\mu_\nu,a_\nu}\Bigg\},\\
    \end{aligned}
\end{equation}
%
\begin{equation}
\label{C-coefficients_nucleon}
\begin{aligned}
     &C^N_V=\frac12Tr\left[\left(-F_1^N\mathbb{1}+F_Q^NQ^L\right)^2\tilde\rho\right],\\
     &C^N_A=\frac12Tr\left[\left(G^N_A\mathbb{1}\right)^2\tilde\rho\right],\\
     &C^N_{V\&A}=\frac12Tr\left[\mathbb{1}G^N_A\left(-F_1^N\mathbb{1}+F_Q^NQ^L\right)\tilde\rho\right],\\
     &C^N_M=\frac12Tr\left[\left(\mathbb{1}F^N_2-F^N_MQ^L\right)^2\tilde\rho\right],\\
     &C^N_{V\&M}=\frac12Tr\left[\left(\mathbb{1}F^N_2-F^N_MQ^L\right)\left(-F_1^N\mathbb{1}+F_Q^NQ^L\right)\rho^K\right],\\
     &C^N_{A\&M}=\frac12Tr\left[\left(\mathbb{1}F^N_2-F^N_MQ^L\right)\mathbb{1}G^N_A\tilde\rho\right],\\    
     &Q^L=\frac{2\sqrt{2}\pi\alpha}{G_F}\left(\frac16\langle r_\nu^2\rangle-a_\nu\right), \quad  |\mu_\nu|^2=Tr\left[\mu_\nu^2\tilde\rho\right],\\
     &K_{\mu_\nu,Z^0}=Tr\left[\mu_\nu\kappa^\dagger\right], \quad K_{\mu_\nu,a_\nu}=Tr\left[a_\nu \mu_\nu\kappa^\dagger\right],\\
     &\kappa_{ij}=\frac12(\zeta^x_{ij}+i\zeta^y_{ij}),
\end{aligned}
\end{equation}
where the neutral weak (NC) and electromagnetic (EM) nucleon form factors are determined by the corresponding vertexes
\begin{equation}
\label{NucleonFF}
\begin{aligned}
     \Lambda^{({\rm NC};N)}_\mu(q)&=\gamma_\mu F_1^N(q^2)-\frac{i}{2m_N}\,\sigma_{\mu\nu}q^\nu F_2^N(q^2)-\gamma_\mu\gamma_5 G_A^N(q^2),\\
     \Lambda^{({\rm EM};N)}_\mu(q)&=\gamma_\mu F_Q^N(q^2)-\frac{i}{2m_N}\,\sigma_{\mu\nu}q^\nu F_M^N(q^2).
\end{aligned}
\end{equation}

The formulas for neutrino-electron scattering are obtained using the substitutions $m_N\rightarrow m_e$, $\delta_{ij}F_1^N(q^2)\rightarrow(\hat{g}_V^e)_{ij}=g_V^e\delta_{ij}+U^*_{ei}U_{ej}$, $\delta_{ij}G_A^N(q^2)\rightarrow(\hat{g}_A^e)_{ij}=g_A^e\delta_{ij}+U^*_{ei}U_{ej}$, $F_2^N(q^2)\rightarrow0$ and neglecting the electron anomalous magnetic moment, $F_M^N(q^2)\rightarrow0$. Thus, we get 
\begin{equation}
\label{FullCSe}
    \begin{aligned}
        \tilde{A}(T)&=\frac{G^2_Fm_e}{4\pi^2}\Bigg[\left(C^e_V-2{\rm Re}\,C^e_{V\&A}+C^e_A\right)+\\
        &+\left(C^e_V+2{\rm Re}\,C^e_{V\&A}+C^e_A\right)\left(1-\frac{T}{E_\nu}\right)^2\\
        &+\left(C^e_A-C^e_V\right)\frac{m_eT}{E^2_\nu}\Bigg]+\frac{\alpha^2|\mu_\nu|^2}{2m^2_e}\left(\frac{1}{T}-\frac{1}{E_\nu}\right),\\
        \tilde{B}_s(T)&=-\frac{G_F\alpha}{\pi}\sqrt{\frac{1}{m_eT}\left(1-\frac{T}{E_\nu}-\frac{m_eT}{2E_\nu^2}\right)}\Bigg\{\Bigg[\left(1-\frac{T}{2E_\nu}\right)g_V^e+\\
        &+\frac{T}{2E_\nu}g^e_A\Bigg]\textrm{Im}K_{\mu_\nu,Z^0}+\Bigg[\left(1-\frac{T}{2E_\nu}\right)+\frac{T}{2E_\nu}\Bigg]\textrm{Im}K_{\mu_\nu,W}-\\
        &-\frac{4\sqrt2\pi\alpha}{G_F}\left(1-\frac{T}{2E_\nu}\right)\textrm{Im}K_{\mu_\nu,a_\nu}\Bigg\},\\
        \tilde{B}_c(T)&=\frac{G_F\alpha}{\pi}\sqrt{\frac{1}{m_eT}\left(1-\frac{T}{E_\nu}-\frac{m_eT}{2E_\nu^2}\right)}\Bigg\{\Bigg[\left(1-\frac{T}{2E_\nu}\right)g_V^e+\\
        &+\frac{T}{2E_\nu}g^e_A\Bigg]\textrm{Re}K_{\mu_\nu,Z^0}+\Bigg[\left(1-\frac{T}{2E_\nu}\right)+\frac{T}{2E_\nu}\Bigg]\textrm{Re}K_{\mu_\nu,W}-\\
        &-\frac{4\sqrt2\pi\alpha}{G_F}\left(1-\frac{T}{2E_\nu}\right)\textrm{Re}K_{\mu_\nu,a_\nu}\Bigg\},\\
    \end{aligned}
\end{equation}
and
\begin{equation}
\label{C-coefficients_electron}
\begin{aligned}
     &C^e_V=\frac12Tr\left[\left(\hat g_V^e+Q^L\right)^2\tilde\rho\right],\quad C^e_A=\frac12Tr\left[\left(\hat g_A^e\right)^2\tilde\rho\right],\\
     &C^e_{V\&A}=-\frac12Tr\left[\hat g^e_A\left(\hat g_V^e+Q^L\right)\tilde\rho\right],\\
     &K_{\mu_\nu,W}=Tr\left[(\hat g_V^e-\mathbb{1}g_V^e)\mu_\nu\kappa^\dagger\right].
\end{aligned}
\end{equation}
%







\end{document}